\documentclass[pra,twocolumn,showpacs,amsfonts,amsmath,floatfix]{revtex4}
\usepackage{epsfig,amsmath,graphicx}

\usepackage{graphicx}
\usepackage{dcolumn}
\usepackage{bm}
\usepackage{hyperref}

\begin{document}


\title{Intensity and phase noise correlations in a dual-frequency VECSEL \\operating at telecom wavelength}

\author{Syamsundar De$^{1}$}
\author{Ghaya Baili$^{2}$ }
\author{Sophie Bouchoule$^{3}$ }
\author{Mehdi Alouini$^{4}$ }
\author{Fabien Bretenaker$^{1}$}
\email{Fabien.Bretenaker@u-psud.fr}
 \affiliation {$^1$Laboratoire Aim\'e Cotton, CNRS-ENS Cachan-Universit\'e Paris Sud 11, 91405 Orsay Cedex, France\\$^2$Thales Research and Technology, 91120 Palaiseau, France\\$^3$Laboratoire de Photonique et de Nanostructures, 91460 Marcoussis, France\\$4$Institut de Physique de Rennes, CNRS-Universit\'e de Rennes I, 35042 Rennes, France}

\date{\today}

\begin{abstract}
The amplitude and phase noises of a dual-frequency vertical-external-cavity surface-emitting laser (DF-VECSEL) operating at telecom wavelength are theoretically and experimentally investigated in detail. In particular, the spectral behavior of the correlation between the intensity noises of the two modes of  the DF-VECSEL is measured. Moreover, the correlation between the phase noise of the radio-frequency (RF) beatnote generated by optical mixing of the two laser modes with the intensity noises of the two modes is investigated. All these spectral behaviors of noise correlations are analyzed for two different values of the nonlinear coupling between the laser modes. We find that to describe the spectral behavior of noise correlations between the laser modes, it is of utmost importance to have a precise knowledge about the spectral behavior of the pump noise, which is the dominant source of noise in the frequency range of our interest (10 kHz to 35 MHz). Moreover, it is found that the noise correlation also depends on how the spatially separated laser modes of the DF-VECSEL intercept the noise from a multi-mode fiber-coupled laser diode used for pumping both the laser modes. To this aim, a specific experiment is reported, which aims at measuring the correlations between different spatial regions of the pump beam. The experimental results are in excellent agreement with a theoretical model based on modified rate equations. 
\end{abstract}
\pacs{42.55.Px, 42.60.Mi,42.55.Ah}

\maketitle
\section{Introduction}
\label{sec:Intro}
A coherent source emitting two optical frequencies with a difference tunable over a wide range and having a high degree of correlation between their fluctuations could find applications in domains such as ultra-stable atomic clocks \cite{Knappe2004,Vanier2005}, pump-probe experiments, generation of high-purity optically-carried radio frequency (RF) signals useful for microwave photonics \cite{Alouini2001,Tonda2006}, and metrology \cite{Nerin1997}. The dual-frequency laser, sustaining two orthogonal linear polarizations with a tunable frequency difference can perfectly fit the above mentioned applications. Such dual-frequency operation has been achieved for different solid-state lasers \cite{Brunel1997, Czarny2004}, but these lasers suffer from relatively strong intensity noise due to relaxation oscillations inherent to their class-B dynamical behavior \cite{Arecchi1984}. The dual-frequency vertical-external-cavity surface-emitting laser (DF-VECSEL) exhibits low noise due to its relaxation oscillation free class-A dynamical behavior, as the photon lifetime ($ \sim $10 ns) inside the cm-long external cavity  is longer than the carriers' lifetime ($ \sim $3 ns) of the semiconductor gain medium. More importantly, the intensity and phase fluctuations of the two polarizations are expected to be strongly correlated as the two laser modes are oscillating inside the same cavity and pumped by the same laser source. Such DF-VECSELs have been demonstrated for several different wavelengths such as 1 $ \mu $m \cite{Baili2009}, 852 nm \cite{Camargo2012}, and recently we have realized DF-VECSEL operating at 1.55 $ \mu $m wavelength \cite{De2014}. The noise properties of such DF-VECSEL are not easy to analyze as the two laser modes are coupled and, in addition to that, phase and intensity fluctuations are linked with each other due to the large Henry factor ($\alpha$-factor) of the semiconductor gain medium \cite{Henry1982}. Previously, we have investigated, both experimentally and theoretically, the spectral behavior of the intensity noises and their correlation \cite{De2013}, and the phase noise spectra of the RF beatnote for different nonlinear coupling  between the two modes of an optically pumped dual-frequency VECSEL operating at 1 $ \mu $m \cite{DeJLT2014}. The aim of the present paper is to report the spectral behavior of intensity noise correlations and phase noise spectra of the RF beatnote for different values of the nonlinear coupling between the two modes for an optically pumped DF-VECSEL  operating at 1.55 $ \mu $m. Moreover, the spectral behavior of the correlations between the phase noise of the RF beatnote and the intensity noises of the two laser modes are also investigated for different coupling strengths between the laser modes. To put our understanding about the noise properties of DF-VCSEL on more solid basis, we investigate the spectral behavior of the pump noise, which is found to be the dominant source of noise within the frequency range of our interest (10 kHz to 35 MHz). Additionally, we figure out how the two laser modes, which are spatially separated inside the active medium but pumped by the same pump laser, are intercepting intensity fluctuations of the pump laser differently, depending on their spatial separation. We prove that this dependence of pump noise properties on the spatial separation between the modes, defining the nonlinear coupling between them, plays an important role in the strength and phase of the correlations between different noises of the two laser modes. All the experimental results are compared with the predictions from a simple theoretical model based on rate equations, as mentioned in Ref.\ \cite{De2013,DeJLT2014}. This permits to probe the generality of this simple theoretical model. 

In Sec.\ \ref{sec:Theory}, we develop the theoretical model to describe the effects of the pump noise determining the spectral behavior of intensity noises, phase noise of the RF beatnote and correlation between all these noises for our DF-VECSEL. We show both the experimental and theoretical results in Sec\ \ref{sec:Results}.

\section{Theoretical Model} 
\label{sec:Theory}
\begin{figure}[htbp]
\centering
\fbox{\includegraphics[width=0.8\columnwidth]{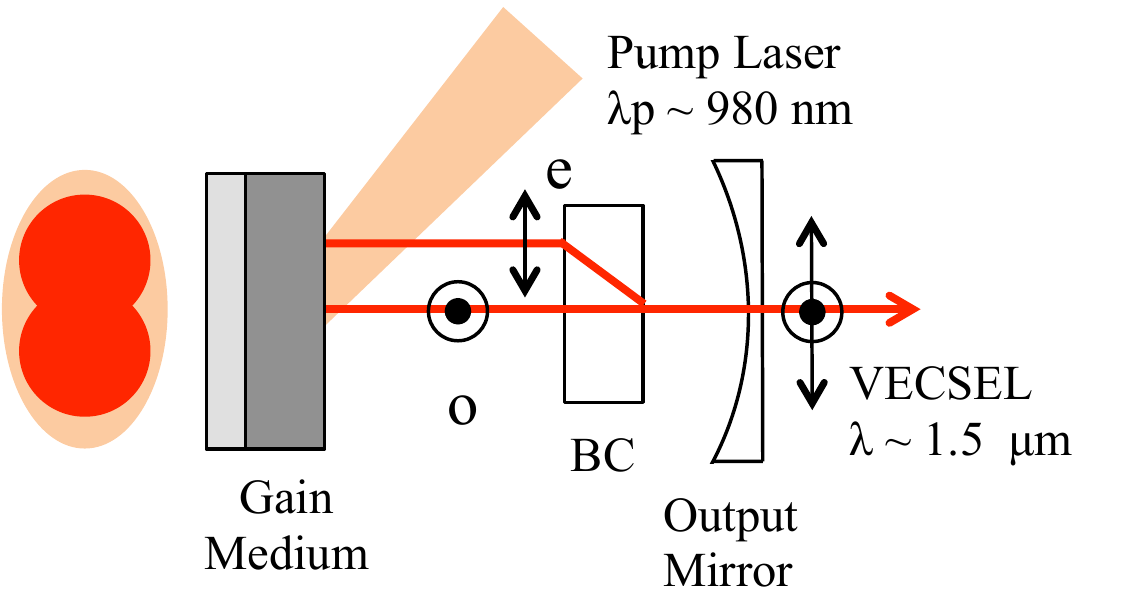}}
\caption{(Color online) Schematic representation of the basic principle of the dual-frequency VECSEL operating at 1.5 $\mu$m. The ordinary (o) and extraordinary (e) eigenpolarizations are spatially separated inside the gain medium to reduce nonlinear coupling. The pumping is done by a diode laser at 980 nm. The picture on the left represents the two partially separated modes in the pumped region of the gain medium.}
\label{fig:1}
\end{figure}
The schematic of the working principle of the DF-VECSEL sustaining simultaneous oscillation of two linear-orthogonal polarizations, which are partially spatially separated in the active medium by an intra-cavity birefringent crystal (BC), is shown in Fig.\ \ref{fig:1}. This partial spatial separation is required in order to make the simultaneous dual-mode operation robust. The DF-VECSEL can be modeled using a more rigorous approach by taking into account the spin-dependent dynamics of the carriers inside the active medium \cite{DePRA2014}. But, here, we follow the approach of Refs.\ \cite{De2013,DeJLT2014} to model the DF-VECSEL using simple rate equations. We indeed know from Ref. \cite{DePRA2014} that this simpler model is good enough to explain the noise properties of the DF-VECSEL and can be derived from the spin-flip model. 

\subsection{Definition of noise spectra and correlations}
The DF-VECSEL schematized in Fig.\ \ref{fig:1} emits both $x-$ (o$ - $) and $y-$ (e$ - $) polarized modes. These modes are characterized by their respective intensities, phases, and frequencies, with the associated noises. Let us thus call $F_x(t)$ and $F_y(t)$ the (dimensionless) number of photons in the $x-$ and $y-$polarized modes, respectively. When these two modes oscillate, their average steady-state intensities are called $F_{x0}$ and $F_{y0}$. We can then write their fluctuating numbers of photons as:
\begin{eqnarray}
F_x(t)&=&F_{x0}+\delta F_x(t)\ ,\label{eqfb01}\\
F_y(t)&=&F_{y0}+\delta F_y(t)\ .\label{eqfb02}
\end{eqnarray}
The relative intensity noise spectra of the laser modes are then defined by:
\begin{eqnarray}
RIN_{x}(f)&=&\frac{\langle \vert \widetilde{\delta F}_{x}(f)\vert^{2}\rangle}{{F_{x0}}^{2}}\ ,\label{eq:16}\\
RIN_{y}(f)&=&\frac{\langle \vert \widetilde{\delta F}_{y}(f)\vert^{2}\rangle}{{F_{y0}}^{2}} ,\label{eq:17}
\end{eqnarray}
where $f$ is the noise frequency and the tilde $\widetilde{\ }$ denotes Fourier-transformed quantities.We also define the normalized correlation spectrum between the intensity noises of the two laser modes as 
\begin{equation}
\Theta_{Fx-Fy}(f)=\frac{\langle\widetilde{\delta F}_{x}(f)\widetilde{\delta F}^{*}_{y}(f)\rangle }{\sqrt{ \langle \vert \widetilde{\delta F}_{x}(f)\vert^{2}\rangle \langle \vert \widetilde{\delta F}_{y}(f)\vert^{2}\rangle}} .\label{eq:28}
\end{equation}

Similarly, we introduce the phase noises $\delta\phi_x(t)$ and $\delta\phi_y(t)$ of the two modes, and the phase noise of the generated beat note:
\begin{equation}
\delta\phi_{Beat}(t)=\delta\phi_x(t)-\delta\phi_y(t)\ .\label{eqfb03}
\end{equation}
In the following, we calculate and measure the normalized correlation spectra between the RF beatnote phase noise and the intensity noises of $x-$ and $y-$polarized modes  respectively defined as
\begin{equation}
\label{eq:29}
\Theta_{Beat-Fx}(f)=\frac{\langle\widetilde{\delta \phi}_{Beat}(f)\widetilde{\delta F}^{*}_{x}(f)\rangle }{\sqrt{ \langle \vert \widetilde{\delta  \phi}_{Beat}(f)\vert^{2}\rangle \langle \vert \widetilde{\delta F}_{x}(f)\vert^{2}\rangle}}\ ,
\end{equation}
and
 \begin{equation}
\label{eq:30}
\Theta_{Beat-Fy}(f)=\frac{\langle\widetilde{\delta \phi}_{Beat}(f)\widetilde{\delta F}^{*}_{y}(f)\rangle }{\sqrt{ \langle \vert \widetilde{\delta  \phi}_{Beat}(f)\vert^{2}\rangle \langle \vert \widetilde{\delta F}_{y}(f)\vert^{2}\rangle}}\ .
\end{equation}

The correlation spectra given by Eqs.\,(\ref{eq:28}), (\ref{eq:29}), and (\ref{eq:30}) are complex quantities with a modulus equal/smaller than 1. A modulus equal to 1 means perfect correlation and the argument gives the phase of the correlation. The next parts of this section are devoted to the derivation of these spectra.
\subsection{Laser rate equations}
Before introducing the noise sources, let us present the rate equations we use for the photon numbers and population inversion numbers of the DF-VECSEL \cite{De2013}:
\begin{eqnarray}
\frac{dF_{x}(t)}{dt}=-\frac{F_{x}(t)}{\tau_{x}}+\kappa N_{x}(t) F_{x}(t)\ ,\label{eq:1}\\
\frac{dF_{y}(t)}{dt}=-\frac{F_{y}(t)}{\tau_{y}}+\kappa N_{y}(t) F_{y}(t)\ ,\label{eq:2}
\end{eqnarray}
\begin{align}
\frac{dN_{x}(t)}{dt}=\dfrac{1}{\tau}[N_{0x}(t)-N_{x}(t)]-\kappa N_{x}(t) [F_{x}(t)+\xi_{xy}F_{y}(t)]\ ,\label{eq:3}\\
\frac{dN_{y}(t)}{dt}=\dfrac{1}{\tau}[N_{0y}(t)-N_{y}(t)]-\kappa N_{y}(t) [F_{y}(t)+\xi_{yx}F_{x}(t)]\ .\label{eq:4}
\end{align}
$ N_{x} $ and $ N_{y} $ are the population inversion numbers for the two modes. $ \tau_{x} $, $ \tau_{y} $ refer to the photon lifetimes inside the cavity for the $x-$ and $y-$polarizations, respectively. $ \tau $ is the population inversion lifetime. $ N_{0x}/\tau $ and $ N_{0y}/\tau$ are the two pumping terms. We allow them to depend on time in order to introduce the pumping noise.  $\kappa $ is proportional to the stimulated emission cross-section. 
The coefficients $\xi_{xy} $ and $ \xi_{yx} $ are the ratios of the cross- to self-saturation coefficients, which also take into account the effect of partial overlap between the two modes inside the gain medium. Therefore, 
\begin{equation}
\label{eq:5}
C=\xi_{xy}\xi_{yx}
\end{equation}
defines the nonlinear coupling constant \cite{Lamb1989}. It must be noted that using the nonlinear coupling constant as defined by Lamb is a convenient way to phenomenologically take into account all the coupling/uncoupling mechanisms independently of their origin \cite{Alouini2000} including mode polarization and geometrical overlap. Of course, in the following we consider only situations for which $C<1$, which is the condition for stable simultaneous oscillation of the two modes to be possible \cite{Lamb1989}.

In the absence of noise, $ N_{0x}$ and $ N_{0y}$ can be replaced by their time averaged values $ \overline{N}_{0x} $ and $  \overline{N}_{0y} $ and the steady-state solutions of Eqs.\ (\ref{eq:1}-\ref{eq:4}) for simultaneous oscillation of the two cross-polarized modes are
\begin{eqnarray}
\label{eq:6}
F_{x0}&=&\frac{(r_{x}-1)-\xi_{xy}(r_{y}-1)}{\kappa\tau(1-C)}\ ,\\
\label{eq:7}
F_{y0}&=&\frac{(r_{y}-1)-\xi_{yx}(r_{x}-1)}{\kappa\tau(1-C)}\ ,\\
\label{eq:8}
 N_{xth}&=&\frac{1}{\kappa\tau_{x}}\ ,\\
\label{eq:9}
N_{yth}&=&\frac{1}{\kappa\tau_{y}}\ .
\end{eqnarray}
Here, $ r_{x}= \overline{N}_{0x}/N_{xth} $ and $ r_{y}=\overline{N}_{0y}/N_{yth} $ define the excitation ratios for the two modes. 

\subsection{Introduction of pump noise}
According to our previous and current experimental observations, the dominant source of noise within the frequency range of 10 kHz to 35 MHz is the pump intensity noise for our DF-VECSEL. To describe how this noise enters  the two laser modes, which are partially spatially separated on the gain structure (see Fig.\ \ref{fig:1}), we model the fluctuations of the two pumping terms as two identical white noises within the considered frequency range (10 kHz to 35 MHz):
\begin{equation}
\label{eq:25}
\langle \vert \widetilde{\delta N}_{0x}(f) \vert ^{2}\rangle=\langle \vert \widetilde{\delta N}_{0y}(f) \vert ^{2}\rangle=\langle \vert \widetilde{\delta N}_{0} \vert ^{2}\rangle\ ,
\end{equation}
with the following correlation spectrum:
\begin{equation}
\label{eq:26}
\langle \widetilde{\delta N}_{0x}(f) \widetilde{\delta N}^{*}_{0y}(f)\rangle=\eta \langle \vert \widetilde{\delta N}_{0} \vert ^{2}\rangle e^{i\psi}\ ,
\end{equation}
where $ \eta $ and  $ \psi $ respectively define the amplitude and phase of correlation, supposed to be constant in the considered frequency range of 10 kHz to 35 MHz. 
In this frequency range, the RIN spectra of the pump noises can be defined as 
\begin{equation}
\label{eq:27}
RIN_{P}=\frac{\langle \vert \widetilde{\delta N}_{0x}(f)\vert^{2}\rangle}{\overline{N}_{0x}^{2}}=\frac{\langle \vert \widetilde{\delta N}_{0y}(f)\vert^{2}\rangle}{\overline{N}_{0y}^{2}} .
\end{equation}
It is important to mention that this way of linking the fluctuations of the population inversion to the pump RIN would no longer be true at frequencies higher than the inverse of electron relaxation time constants, i. e. higher than several hundreds of MHz. Moreover, in the case of an electrically pumped VECSEL, the same formalism could be used to introduce the noise in the injection current \cite{Roy1988}.

In the following, we predict how these pump noise properties allow us to calculate the actual correlation spectra between the different noises of the DF-VECSEL. 

\subsection{Intensity noises}
The impact of these pump noises on the laser intensity noise is derived by performing standard linearization of Eqs.\ (\ref{eq:1}-\ref{eq:4}) around the steady-state solutions (\ref{eq:6}-\ref{eq:9}). After simplification, we obtain the following expression relating the photon number fluctuations $[\widetilde{\delta F}_{x}(f)$, $\widetilde{\delta F}_{y}(f)]$ of the two laser modes to the two pump fluctuations $[\widetilde{\delta N}_{0x}(f) $, $ \widetilde{\delta N}_{0y}(f)]$:
\begin{equation}
\label{eq:10}
\begin{bmatrix}
\widetilde{\delta F}_{x}(f)\\
\widetilde{\delta F}_{y}(f)
\end{bmatrix}
=
\begin{bmatrix}
M_{xx}(f) & M_{xy}(f)\\
M_{yx}(f) & M_{yy}(f)
\end{bmatrix}
\begin{bmatrix}
\widetilde{\delta N}_{0x}(f)\\
\widetilde{\delta N}_{0y}(f)
\end{bmatrix} ,
\end{equation}
with
\begin{eqnarray}
\label{eq:11}
M_{xx}(f)&=&\frac{1}{\tau}\frac{[\frac{1}{\tau_{y}}-\frac{2i\pi f}{\kappa F_{y0}}(r_{y}/\tau - 2i\pi f)]}{\Delta (f)}\ ,\\
\label{eq:12}
 M_{xy}(f)&=&-\frac{\xi_{xy}}{\tau \tau_{x} \Delta (f)}\ ,\\
\label{eq:13}
 M_{yx}(f)&=&-\frac{\xi_{yx}}{\tau \tau_{y} \Delta (f)}\ ,\\
\label{eq:14}
M_{yy}(f)&=&\frac{1}{\tau}\frac{[\frac{1}{\tau_{x}}-\frac{2i\pi f}{\kappa F_{x0}}(r_{x}/\tau - 2i\pi f)]}{\Delta (f)}\ ,
\end{eqnarray}
and

\begin{align}
\label{eq:15}
\Delta (f)=& [\frac{1}{\tau_{x}}-\frac{2i\pi f}{\kappa F_{x0}}(r_{x}/\tau - 2i\pi f)]
\nonumber \\\times&[\frac{1}{\tau_{y}}-\frac{2i\pi f}{\kappa F_{y0}}(r_{y}/\tau - 2i\pi f)] 
-C/\tau_{x}\tau_{y}.
\end{align}
The RIN spectra of the two laser modes and the intensity noise correlations can then be obtained from Eqs.\,(\ref{eq:16}-\ref{eq:28}).

\subsection{Phase noise of the RF beatnote}
The RF beatnote is generated by optically mixing the two orthogonal linear polarizations of the laser. The phase noise of this beat note is thus the difference between the two optical phase noises of the two modes, as shown by Eq.\,(\ref{eqfb03}). We consider two different mechanisms \cite{DeJLT2014}, described in the following, by which the pump intensity noise can induce such phase fluctuations.
\subsubsection{Effect of the large Henry factor}
The first mechanism involves the phase-intensity coupling present in semiconductor active media, characterized by the Henry factor $\alpha$ \cite{Henry1982}. The effect of this coupling is that the population inversion also drives the phase of the laser, leading to
\begin{equation}
\label{eq:18}
\frac{d\phi_H(t)}{dt}=\dfrac{d\phi_{x}(t)}{dt}-\dfrac{d\phi_{y}(t)}{dt}=\frac{\alpha}{2} \kappa [N_{x}(t)-N_{y}(t)] .
\end{equation}
Here $ \phi_H(t) $ denotes the phase noise of the beatnote due to this mechanism. The Fourier transform $ \widetilde{\delta \phi_H}(f) $ of this RF phase noise can be readily obtained thanks to Eqs.\ (\ref{eq:1}-\ref{eq:4}) as
\begin{equation}
\label{eq:19}
\widetilde{\delta \phi}_H(f)=\frac{\alpha}{2}[\frac{\widetilde{\delta F}_{x}(f)}{F_{x0}}-\frac{\widetilde{\delta F}_{y}(f)}{F_{y0}}] .
\end{equation}
One can notice that $ \widetilde{\delta \phi}_H(f) $ can vanish if the two terms inside the square brackets in the right hand side of Eq.\ (\ref{eq:19}) are equal, i.e., if the intensities of the two modes are balanced and their noises are perfectly correlated. In our DF-VECSEL, the phase-intensity coupling effect is strong  due to the large value of $ \alpha $ in the QW-based active medium. Moreover, the two intensity noises are never perfectly correlated. This can give rise to a strong phase noise for the RF beatnote, as we will see below. 

\subsubsection{Pump noise induced thermal fluctuations}
In addition to the phase-intensity coupling mechanism, pump noise induced thermal fluctuations of the refractive index of the active medium have been found to be another dominant source of phase fluctuations degrading the noise performance of our DF-VECSEL. As derived in Ref.\ \cite{DeJLT2014}, the corresponding RF phase fluctuation is given in the frequency domain by
\begin{align}
\label{eq:20}
\widetilde{\delta \phi}_T(f)=\frac{c}{i \lambda_0 f} \Gamma_T R_T  H(f) \frac{2 P_p}{\overline{N}_{0x} + \overline{N}_{0y}}\nonumber\\
 \times(\widetilde{\delta N}_{0x}(f) -\widetilde{\delta N}_{0y}(f))\ .
\end{align}
Here, $\lambda_0 $ is the central wavelength for the two lasing modes, $R_T $ is the thermal impedance of the gain structure, and $P_p$ the total pump power. $ \Gamma_T $ holds for the refractive index variation with temperature, given by
\begin{equation}
\label{eq:21}
\Gamma_{T}=\dfrac{L_{SC}}{L_{ext}}\frac{d \overline {n}}{d T} ,
\end{equation}
where $ {d \overline {n}}/{d T} $ denotes the thermal sensitivity of the average refractive index in the structure, and $ L_{SC} $ and $ L_{ext} $ are the lengths of the semiconductor structure and the external cavity, respectively. $H(f)$ stands for the transfer function of the thermal response of the structure, modeled by a simple first order filter:
\begin{equation}
\label{eq:22}
\vert H(f)\vert^{2}=\frac{1}{1+(2\pi f \tau_{T})^{2}}\ .
\end{equation}
The bandwidth of this filtering effect is determined by the thermal response time  $ \tau_{T} $ of the semiconductor structure, which is approximately given by \cite{Laurian2010}:
\begin{equation}
\label{eq:23}
\tau_{T} \simeq \frac{w_{p}^{2}}{2 \pi D_{T}}\ ,
\end{equation}
where $ w_p $ is the waist of the pump beam on the semiconductor structure and $ D_{T} $ is the thermal diffusion coefficient. 

\subsubsection{Total RF phase noise}
The power spectral density (PSD) of the phase noise of the RF beatnote is given by the sum of the two contributions derived above:
\begin{align}
\label{eq:24}
\vert\widetilde{\delta \phi}_{Beat}(f) \vert^{2}&=\vert(\widetilde{\delta \phi}_{H}(f)+\widetilde{\delta \phi}_{T}(f) )\vert^{2}\nonumber\\
&=\vert\widetilde{\delta \phi}_{H}(f) \vert^{2}+\vert\widetilde{\delta \phi}_{T}(f) \vert^{2}\nonumber\\
&+ 2 \mathrm{Re}[{\widetilde{\delta \phi}_{H}(f)\widetilde{\delta \phi}_{T}(f)^*}]\ .
\end{align}
The important point to be mentioned here is that to obtain the total phase noise of the RF beatnote, the two above mentioned mechanisms, $ \alpha $-factor effect and thermal effect, should be added coherently since these two mechanisms are driven by the same source, i.e. the pump intensity noise. This is an important modification of the model compared to  Ref.\ \cite{DeJLT2014}, where the total phase noise had been obtained by adding the PSDs corresponding to these two mechanisms, hence the last term in the right hand side of Eq.\ (\ref{eq:24}) had been neglected.

\begin{figure*}[]
\centering
\includegraphics[width=1.9\columnwidth]{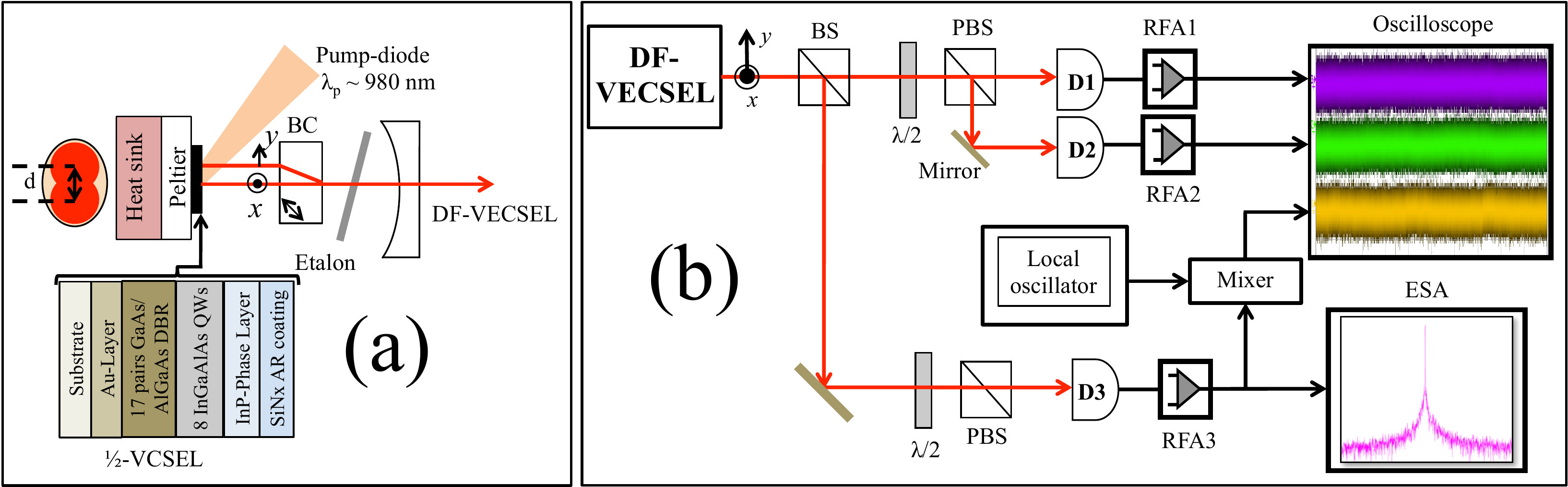}
\caption{(Color online) Schematic of noise correlation measurement setup. (a) Architecture of the DF-VECSEL cavity, $d$: spatial separation between the two eigenpolarizations. (b) Noise measurement setup. $ \lambda/2 $: half-wave plate, BS: beam spliter, PBS: polarization beam spliter, D: detectors, RFA: RF amplifiers, ESA: electrical spectrum analyzer. }
\label{fig:2}
\end{figure*}
\section{Results: experimental and theoretical}
\label{sec:Results}
\subsection{Noise correlation measurement setup}
The experimental setup is depicted in Fig.\ \ref{fig:2}. 
Figure\ \ref{fig:2}(a) shows the architecture of the cavity of the DF-VECSEL operating at 1.55 $ \mu $m. It is based on a 1/2-VCSEL structure
including an InP/InGaAlAs multi-quantum-well active region, and a GaAs/AlAs Bragg mirror \cite{Tourrenc2008}. The substrate used is polycrystalline chemical vapor deposition (CVD) diamond \cite{Zhao2011}. An anti-reflection (AR) coating at the pump wavelength is deposited on the top surface, and the resonant mode of the 1/2-VCSEL microcavity is adjusted to coincide
with the gain maximum \cite{Zhao2011}.

%

The top InP layer, acting as a phase layer, is etched in such a way that the position  of the resonant mode of the 1/2-VCSEL microcavity nearly coincides with the gain maximum after deposition of the AR layer. The temperature of the 1/2-VCSEL structure is maintained at 20$ ^\circ $C using a Peltier thermoelectric cooler. The gain structure is optically pumped, with an incidence angle of about 40$ ^\circ $, by a continuous wave (CW) multi-mode fiber coupled 980 nm diode laser delivering up to 4~W optical power. The length of the cavity is equal to 4.76 cm, giving a free spectral range (FSR) of 3.15 GHz. The output coupler of the cavity is a concave mirror with 99.4\% reflectivity at 1.55 $ \mu $m and a radius of curvature of 5~cm. This cavity configuration leads to laser mode waists of 72 $\mu$m on the gain structure. The pump spot size on the gain structure is adjusted to have maximum and nearly identical power for the two polarizations. In this configuration, the intracavity photon lifetime is of the order of 10~ns, which is larger than the carrier lifetime (typically few hundreds of ps above lasing threshold) in the 1/2-VCSEL structure. This ensures relaxation oscillation free class-A dynamical behavior of our DF-VECSEL.    
Simultaneous oscillation of the two polarizations is achieved by spatially separating them inside the gain medium using an intra-cavity birefringent crystal (BC), which reduces the nonlinear coupling  between the modes below unity \cite{De2014}. A 150-$\mu$m thick uncoated glass etalon inserted between the intracavity BC and the output coupler of the cavity forces each polarization to be longitudinally monomode. 
 
Figure\ \ref{fig:2}(b) shows the schematics of the noise correlation measurement setup.  The noise correlations have been measured for two values of the nonlinear coupling $C$ (see Eq.\,(\ref{eq:5})). These two different values are achieved by using 1 mm- and 0.2 mm-thick BCs, which respectively correspond to spatial separations $d$ of $100\,  \mu $m and $20 \, \mu $m between the modes inside the gain structure. These respectively lead to weak ($ C\simeq 0.11$)  and moderately strong ($ C\simeq 0.74$) nonlinear coupling between the two modes. Of course, both these values must remain below 1 if one wants to achieve simultaneous oscillation of the two modes. All the measurements have been done for a pump power of 1.7~W, which leads to total laser powers of 25 mW and 60 mW for the 1 mm- and 0.2 mm-thick BCs, respectively. The output beam of our DF-VECSEL is divided into two parts using a beam splitter (BS), and then one part is used for intensity noise measurement, whereas the other one is utilized to generate the beatnote and measure its phase noise. To measure the intensity noises of the two polarizations, we separate them using the combination of  a half-wave plate ($ \lambda/2 $) followed by a polarization beam spliter (PBS). Thereafter, the two modes are detected using photodiodes (D1, D2), then amplified by two identical RF amplifiers (RFA1, RFA2), and finally sent to the two channels of a high purity digital oscilloscope. To obtain the beatnote, the two polarizations are mixed using the combination of a $ \lambda/2 $-plate in front a PBS. The RF beat signal is detected using a high-speed photodiode (D3), then amplified using a RF amplifier (RFA3), and finally the spectrum of the beatnote is recorded using an electrical spectral analyzer (ESA). To measure the phase noise of the RF beatnote, it is down shifted to intermediate frequency (IF) by mixing it with a high purity local oscillator. Thereafter, the IF signal is sent to the third channel of the oscilloscope. All the channels are recorded simultaneously and then the oscilloscope data are processed to obtain the noises and their correlation spectra.

\subsection{Pump noise measurement}\label{sec:ResultsPump}
\begin{figure}[]
\centering
\includegraphics[width=0.8\columnwidth]{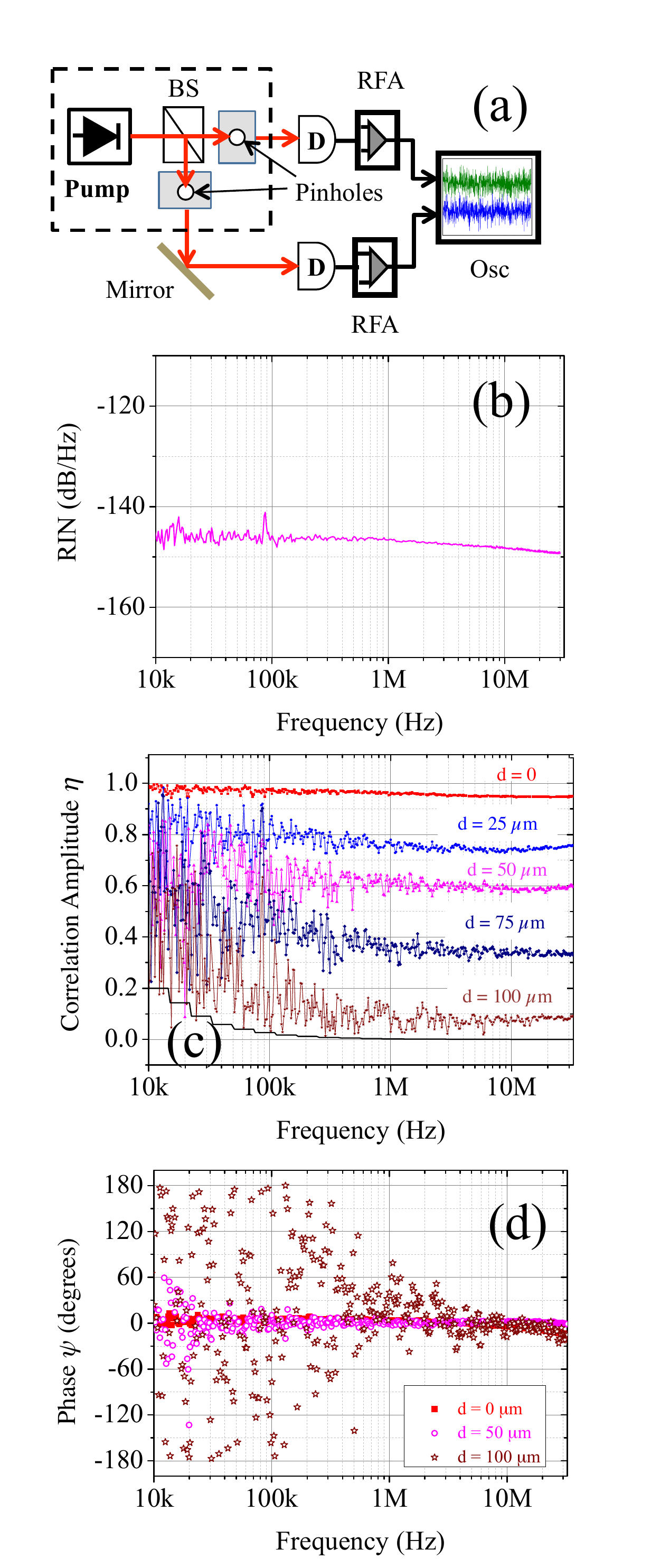}
\caption{(Color online) (a) Schematic of the pump noise correlation measurement setup. BS: beam spliter D: detectors, RFA: RF amplifiers. (b) Pump RIN spectrum. (c) Noise correlation amplitude spectra for different values of $d$. The black thin curve (at the bottom) is the measurement floor. (d) Noise correlation phase spectra for three values of $d$. The phase measurements for $d=25\,\mu\mathrm{m}$ and $d=75\,\mu\mathrm{m}$ are not reproduced here for the sake of clarity, but also give values of $\psi$ which are close to 0. }
\label{fig:3}
\end{figure}

As inferred by our theoretical model in Sec.\ \ref{sec:Theory}, the spectral behaviors of the pump noises entering the two laser modes play an important role to determine the correlation between the different noises of the DF-VECSEL. Here, we experimentally investigate the pump noises and their correlations. In our DF-VECSEL, although the two modes are pumped by one diode laser, their spatial separation inside the gain medium makes them face different regions of the pump beam coming from a multimode fiber [see Fig. \ref{fig:2}(a)]. In order to fully characterize the pump noises seen by the two modes, we mimic this situation by the setup schematized in Fig.\ \ref{fig:3}(a). Here, the pump beam is separated into two paths using a BS. In each path, we introduce a pinhole of identical radii (75 $ \mu $m) mimicking the two laser modes (72 $ \mu $m radius). 

The two pinholes are placed on translation stages, which permit to obtain the situation where one pinhole acts as the exact image of the other. This situation reproduces the condition of perfect overlapping of the two laser modes inside the gain medium, i.e. $ d = 0 $. Therefore, by translating one of the pinholes in the horizontal direction we can reproduce the situations of different spatial separations $d$ between the modes inside the gain structure of our DF-VECSEL. After passing through the pinholes, the pump beams are detected using photodiodes (D), then amplified with identical RFAs, and finally recorded with a digital oscilloscope. Then the oscilloscope data are processed to obtain the spectra of pump noises and their correlation.

We have measured pump noises for different values of $d$ such as 0,  $25\, \mu$m, $50\,\mu $m, $75\,\mu $m, and $100\,\mu $m. In all these cases, the RIN spectra of the two pump beams are nearly identical and given by Fig.\ \ref{fig:3}(b). This confirms that the pump noises entering into the two spatially separated laser modes are identical white noises as assumed in Eq.\ (\ref{eq:25}). Figures \ref{fig:3}(c) and \ref{fig:3}(d) respectively show the pump noise correlation amplitude and phase ($ \eta $ and $ \psi $ of Eq. (\ref{eq:26}), respectively) spectra for different values of $d$. Figure\ \ref{fig:3}(c) shows that the pump noises entering the two spatially separated laser modes are partially correlated ($ \eta < 1 $ ) and the degree of correlation decreases with the increase of spatial separation $ d $ between the modes.  Moreover, the correlated part of the pump intensity noises entering into the two laser modes are in phase ($ \psi = 0 $) for all values of $d$,  as shown in Fig.\ \ref{fig:3}(d). Besides, these results confirm our assumptions, that $\eta$ and $\psi$ do  not depend on the noise frequency [see Eq.(\ref{eq:26})]. The degree of correlation of the pump noises entering into the two laser modes decreases with $d$ since the spatial distribution of pump intensity on the gain structure is not uniform due to the speckle pattern coming from the different modes of the multimode fiber carrying the pump beam. 

\subsection{Intensity noises and their correlations}
This section reports the spectral behavior of the intensity noises of the two polarizations and their correlations for weak ($ C = 0.11 $) and moderately strong ($ C = 0.74 $) couplings, obtained both experimentally and theoretically.
\begin{figure}[htbp]
\centering
\includegraphics[width=1.0\columnwidth]{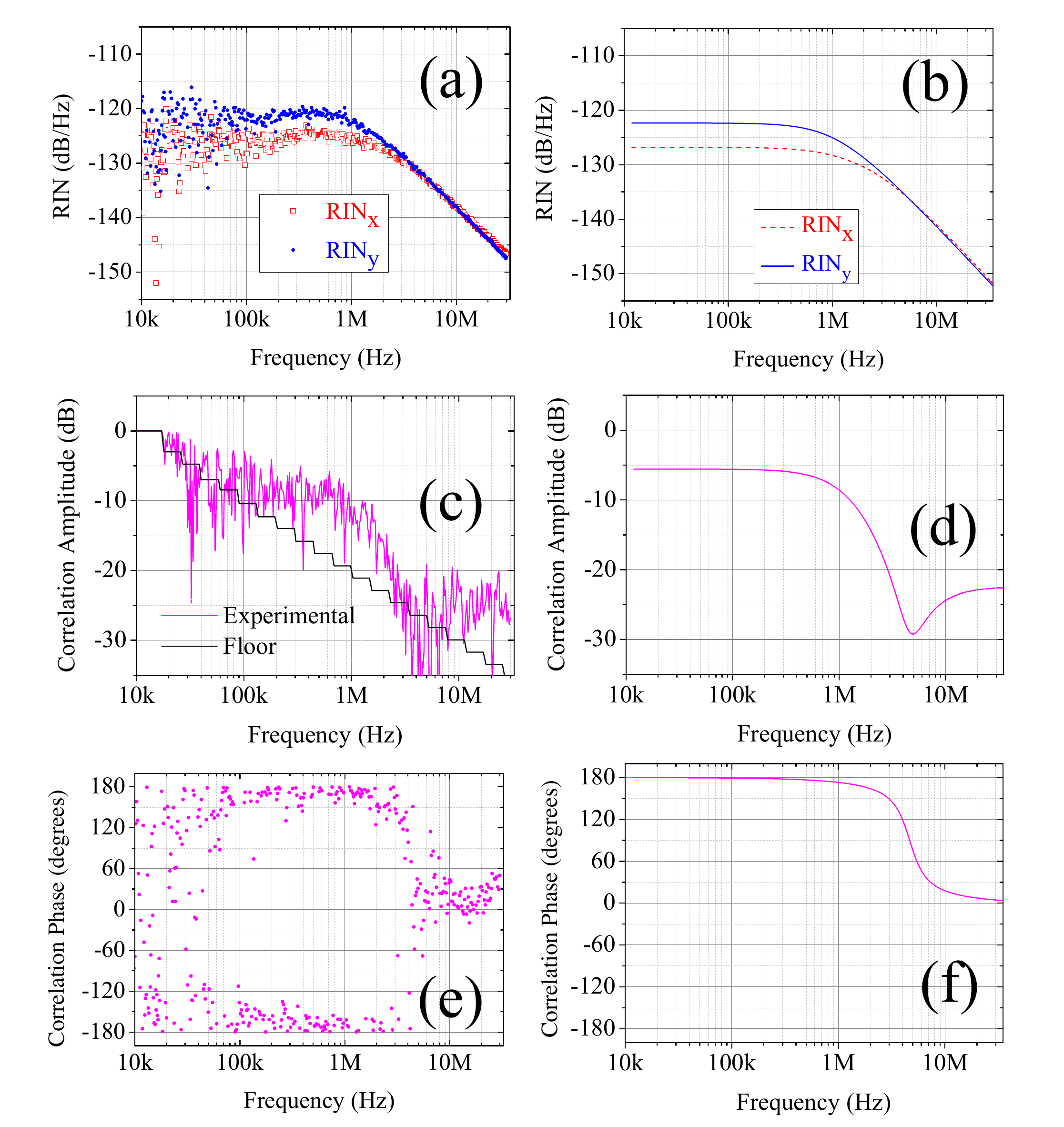}
\caption{(Color online) Results for weak coupling ($ C = 0.11 $): RIN spectra for two polarizations: (a) experimental, (b) theoretical; Correlation amplitude: (c) experimental, (d) theoretical. The black thin line in (c) is the measurement floor; correlation phase: (e) experimental, (d) theoretical. Parameters used for theory: $ r_x = 1.2 $, $ r_y = 1.15 $, $ \tau_x = 9.9 $ ns, $ \tau_y = 10.4 $ ns, $ \tau = 1$ ns, $ \eta = 0.2 $, $ \psi = 0 $, $ RIN_p = -145 $ dB/Hz.}
\label{fig:4}
\end{figure}
The results for weak coupling are shown in Fig.\ \ref{fig:4}. Figures \ref{fig:4}(a),(b) reproduce the RIN spectra of the two modes obtained from experiment and theory, respectively. The RIN spectra for both laser modes are analogous to first-order transfer functions, illustrating the class-A dynamical behavior of our DF-VECSEL. The RINs of the two modes are slightly different as their intra-cavity losses and/or gains are not exactly identical. The experimental and theoretical correlation amplitude spectra are given in Figs.\ \ref{fig:4}(c),(d), respectively.  For weak coupling, the degree of correlation between the intensity noises is nearly about - 7 dB for frequencies lower than 1 MHz. Then, the correlation amplitude drops to a very low value (- 30 dB) around 5 MHz, and then it starts to increase slightly and becomes approximately -25 dB for all frequencies higher than 10 MHz as will be explained hereafter. Figures \ref{fig:4}(e),(f) respectively show the experimental and theoretical correlation phase spectra. These spectra prove that the correlation phase is $ \pi $ for frequencies lower than 1~MHz, then it starts to roll down and reaches zero at about 5 MHz and remains equal to zero for all higher frequencies. 

\begin{figure}[htbp]
\centering
\includegraphics[width=1.0\columnwidth]{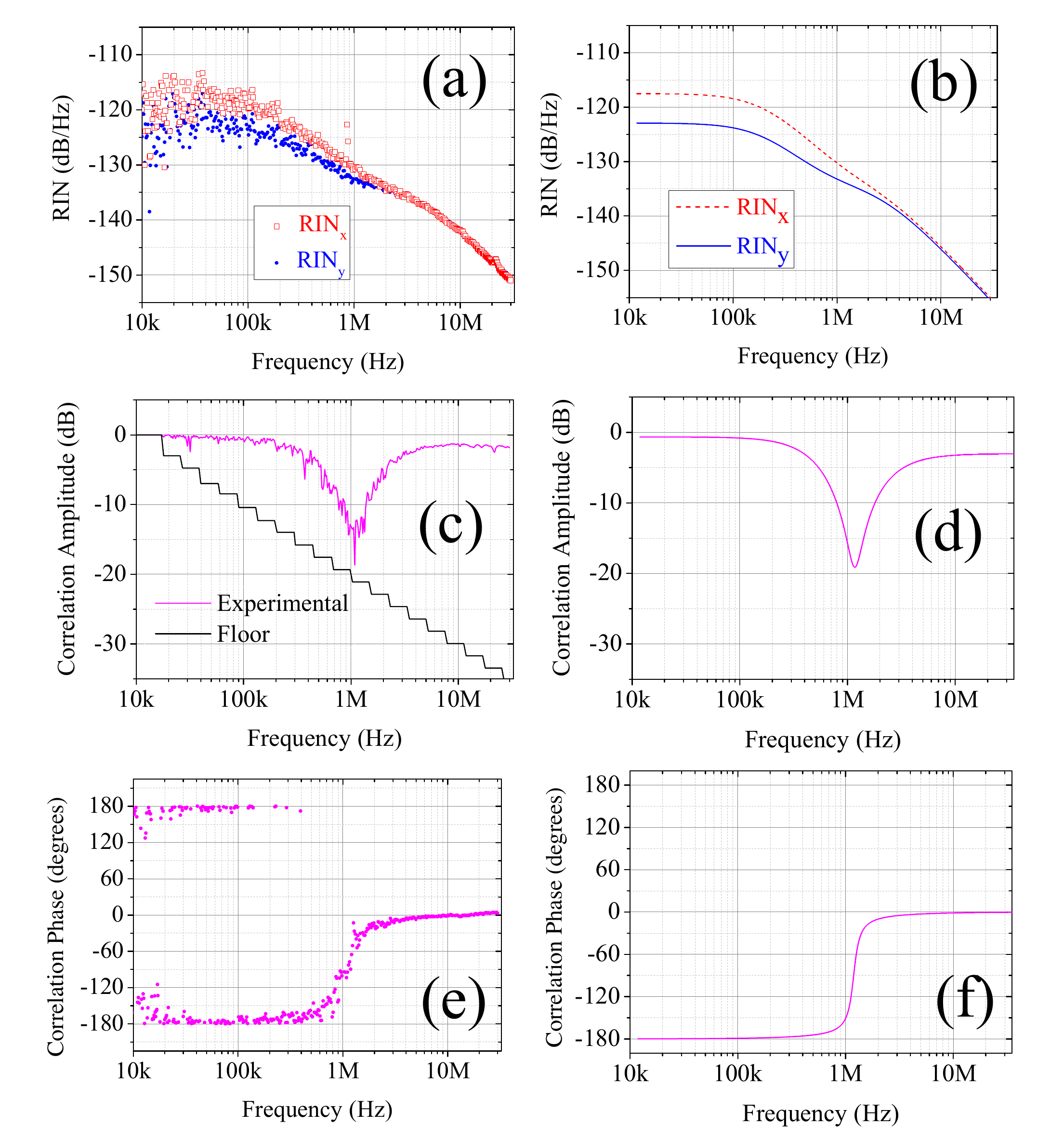}
\caption{(Color online) Results for moderately strong coupling ($ C = 0.74 $) case. RIN spectra for two porizations (a) experimental, (b) theoretical; Correlation amplitude (c) experimental, (d) theoretical; correlation phase (e) experimental (d) theoretical. Parameters used for theory: $ r_x = 1.4 $, $ r_y = 1.42 $, $ \tau_x = 16.8 $ ns, $ \tau_y = 17.6 $ ns, $ \tau = 1$ ns , $ \eta = 0.75 $, $ \psi = 0 $, $ RIN_p = -145 $ dB/Hz.}
\label{fig:5}
\end{figure}

Figure \ref{fig:5} reveals the results for the moderately strong coupling situation. The RIN spectra, both experimental (Fig.\ \ref{fig:5}(a)) and theoretical (Fig.\ \ref{fig:5}(b)), again illustrate the class-A dynamical behavior of our laser as there are no relaxation oscillation peaks in the spectra. The inequality of the RINs for the two modes is coming from the difference of losses and/or gains for them. The notable change for the RIN spectra for moderately strong coupling is the RIN effective cutoff frequency which is now at around 100 kHz as compared to 1 MHz for weak coupling. Moreover one can notice that there is a change of slope around 1 MHz for moderately strong coupling. The spectral behaviors of the correlation amplitude (Figs.\ \ref{fig:5}(c),(d)) and phase (Figs.\ \ref{fig:5}(e),(f)) are also significantly different for the moderately strong coupling compared with the weak one. For moderately strong coupling, correlation amplitude is high (- 2 dB) except from the fact that there is a dip at about 1 MHz (Fig.\ \ref{fig:5}(c),(d)). Moreover, there is a $ \pi $-phase jump in the correlation phase spectra at about 1 MHz, whereas the correlation phase is $ \pi $ for lower frequencies and zero for higher frequencies as depicted in Fig.\ \ref{fig:5}(e),(f). For both weak and moderately strong coupling, the theoretical model nicely agrees with the experiment. The parameter values used for simulations for both coupling condition are obtained from the experimental conditions except from the value of $ \tau $, which is taken from previous experiments \cite{De2014}. Here, the important points to note are the values of pump RIN, pump noise correlation amplitude $ \eta $, and phase $ \psi $ used for different coupling situations. The pump RIN is equal to -145 dB/Hz for all coupling situations, whereas $ \eta = 0.2 $ for weak coupling ($ d = 100 $~$ \mu $m), $ \eta = 0.75 $ for moderately strong coupling ($ d = 20 $~$ \mu $m) and $ \psi $ is zero in both cases. These values are obtained from the pump noise measurements as shown in Sec.\ \ref{sec:ResultsPump}. 

\begin{figure}[htbp]
\centering
\includegraphics[width=1.0\columnwidth]{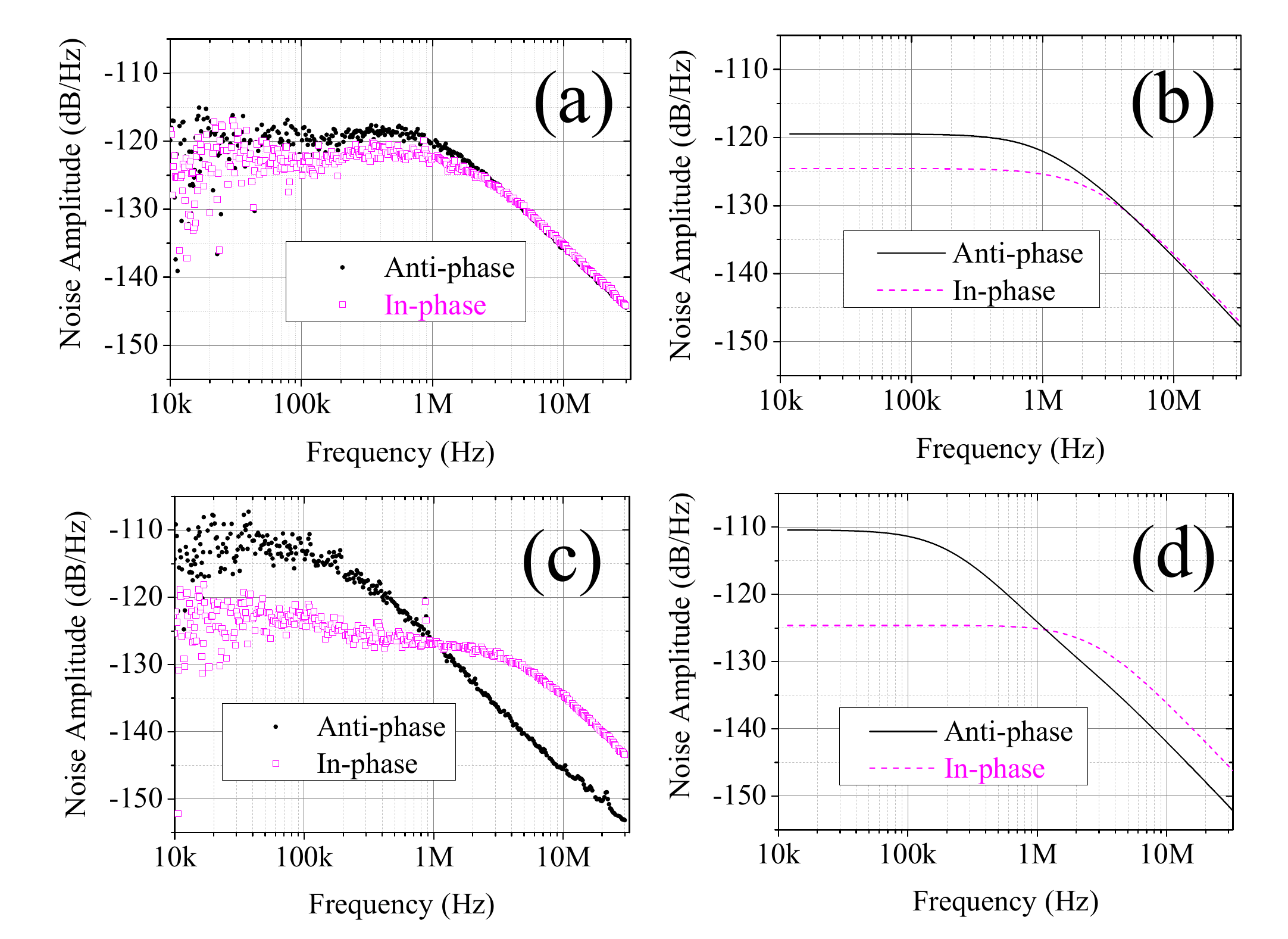}
\caption{(Color online) In-phase and anti-phase noise spectra. Weak coupling: ($ C = 0.11 $) (a) experimental, (b) theoretical; Moderately strong coupling: ($ C = 0.74 $) (c) experimental, (d) theoretical.}
\label{fig:6}
\end{figure}

To gain deeper physical understanding of the spectral behavior of the correlation between the intensity noises of the two modes of our DF-VECSEL and the dependence of noise correlation spectra on coupling strengths, we can invoke the analogy of our DF-VECSEL with a two-coupled mechanical oscillator system. Indeed, the class-A dynamics of the two modes of our DF-VECSEL has some similarities with the over-damped behavior of a two-coupled mechanical oscillator system. Therefore, analogous to any coupled-oscillator system, the dynamical behavior of our DF-VECSEL must rather be analyzed by considering the eigenrelaxation mechanisms of the global system such as in-phase and anti-phase relaxation mechanism \cite{Otsuka1992}. The transfer functions of the in-phase and anti-phase response can be calculated by diagonalizing the $2\times2$ matrix of Eq.\ (\ref{eq:10}). To this aim, the in-phase and anti-phase intensity fluctuations at frequency $ f $ are defined as 
\begin{align}
\widetilde{\delta F}_{\mathrm{In}}(f)&= \widetilde{\delta F}_{x}(f)+\widetilde{\delta F}_{y}(f),\\
\widetilde{\delta F}_{\mathrm{Anti}}(f)&= \widetilde{\delta F}_{x}(f)-\widetilde{\delta F}_{y}(f)
\end{align}
 The normalized in-phase and anti-phase noise spectra for two coupling situations are depicted in Fig.\ \ref{fig:6}. In case of weak coupling, the anti-phase mechanism dominates over the in-phase one for frequencies lower than 5 MHz, whereas for higher frequencies the in-phase mechanism becomes dominant but by a very little margin, as shown by both experiment (Fig.\ \ref{fig:6}(a)) and theory (Fig.\ \ref{fig:6}(b)). This explains the $ \pi $ and zero values of the correlation phase respectively for frequencies lower and higher than 5 MHz in Figs.\ \ref{fig:4}(e),(f). Moreover, the correlation amplitude is low (- 7 dB) for frequencies lower than 5 MHz and it becomes even lower (- 25 dB)  for higher frequencies (Fig.\ \ref{fig:4}(c),(d)). This low noise correlation is related to the fact that the noise sources (pump noises) for the two laser modes are hardly correlated ($ \eta = 0.1 $). Moreover, the difference of correlation amplitudes for frequencies higher and lower than 5 MHz is coming from the different degrees of dominance between anti-phase and in-phase mechanisms. 

For moderately strong coupling, the in-phase and anti-phase noise spectra are represented in Fig.\ \ref{fig:6}(c) (experimental) and Fig.\ \ref{fig:6}(d) (theoretical). These figures show that the anti-phase response strongly dominates over the in-phase one for frequencies lower than 1 MHz, whereas above this cut-off frequency, the in-phase mechanism becomes strongly dominant. As a result, the correlation phase is $ \pi $  for frequencies lower than 1 MHz and zero for higher frequencies, as shown by Figs.\ \ref{fig:5}(e),(f). Moreover, the correlation amplitude is high (- 2 dB) except for the dip around 1 MHz in Figs.\ \ref{fig:5}(c),(d). This is linked to the fact that the coupling between the two modes is stronger and the pump noises (the noise sources) for the two modes are strongly correlated ($ \eta = 0.75 $). The dip in the correlation amplitude occurs as the nearly identical intensity fluctuations of the two modes interfere destructively due to the $ \pi $-phase jump around 1 MHz as shown in Figs.\ \ref{fig:5}(e),(f). Finally, the change of slope in the RIN spectra of the two laser modes at about 1 MHz as in Figs.\ \ref{fig:5}(a),(b) are also linked with the transition from dominant anti-phase to dominant in-phase mechanism around this cut-off frequency. 
  
\subsection{RF beatnote and its phase noise}
\begin{figure}[htbp]
\centering
\includegraphics[width=1.0\columnwidth]{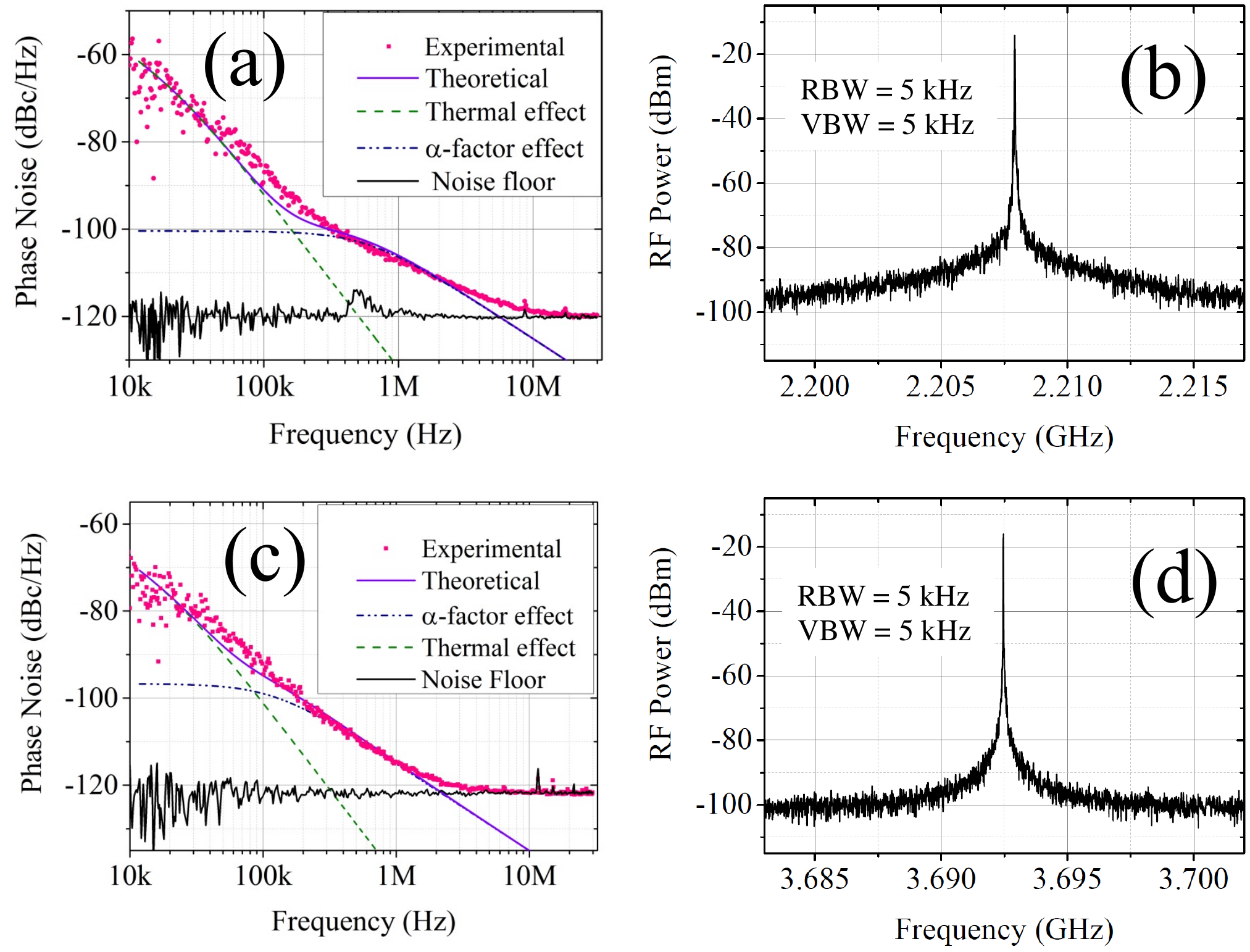}
\caption{(Color online) Results for RF phase noise measurement; weak coupling: (a) phase noise, (b) beatnote, and moderately strong coupling: (c) phase noise, (d) beatnote; RBW: resolution bandwidth, VBW: video bandwidth. Parameters used for simulation: $ \alpha = 10 $, $ D_T = 0.372$~$\mathrm{cm^2s^{-1}}$, $ d \overline{n}/dt=2.0 \times 10^{-4} $~$ \mathrm{K^{-1}} $, $ L_{SC}=5 $~$ \mu$m, $ L_{ext}= 4.76 $~cm, $ w_{p}=50 $~$ \mu $m, $ R_T=12$~$ \mathrm{K.W^{-1}} $, $ P_p = 1.7 $~W.}
\label{fig:7}
\end{figure}

In this section, we analyze the spectral behavior of the phase noise of the RF beatnote for both weak and moderately strong coupling. Figures \ref{fig:7}(a),(b) respectively show the RF phase noise and the corresponding beatnote spectra for the weak coupling case. The beatnote spectrum, recorded with an ESA (Resolution bandwidth (RBW) = video bandwidth (VBW) = 5 kHz), is centered around 2.208 GHz and is sitting on a few MHz wide pedestal (see Fig. \ref{fig:7}(b)). This pedestal is coming from the phase noise whose PSD is shown in Fig.\ \ref{fig:7}(a).
 The phase noise PSD and the corresponding beatnote spectrum for moderately strong coupling situation are reproduced in Figs.\ \ref{fig:7}(c),(d), respectively. For moderately strong coupling also, the beatnote is sitting on a noise pedestal of few MHz width (Fig.\ \ref{fig:7}(d)). The PSD of the phase noise giving rise to the pedestal for the RF beatnote is reported in Fig.\ \ref{fig:7}(c). For both weak and moderately strong coupling, the total theoretical RF phase noise is obtained by coherent addition of the two different mechanisms induced by pump noise: (i)  thermal variation of the refractive index of the active medium [Eq.\ (\ref{eq:20})], which dominates below the cut-off frequency of about 100 kHz (dashed-green line in Figs.\ \ref{fig:7}(a),(c)), and (ii) phase-intensity coupling [Eq.\ (\ref{eq:18})],  dominating for higher frequencies due to the large $ \alpha $-factor of the semiconductor active medium (dash-dotted blue line in Figs.\ \ref{fig:7}(a),(c)). In Figs.\ \ref{fig:7}(a),(c), the pink filled circles represent the measured phase noise PSD, whereas the purple solid line is the total RF phase noise obtained from the theoretical model  [Eq.\ (\ref{eq:24})].  We obtain a very good agreement between theory and experiment for both weak and moderately strong coupling (Figs.\ \ref{fig:7}(a),(c)).

\subsection{Correlation between RF phase noise and intensity noises}
\begin{figure}[htbp]
\centering
\includegraphics[width=1.0\columnwidth]{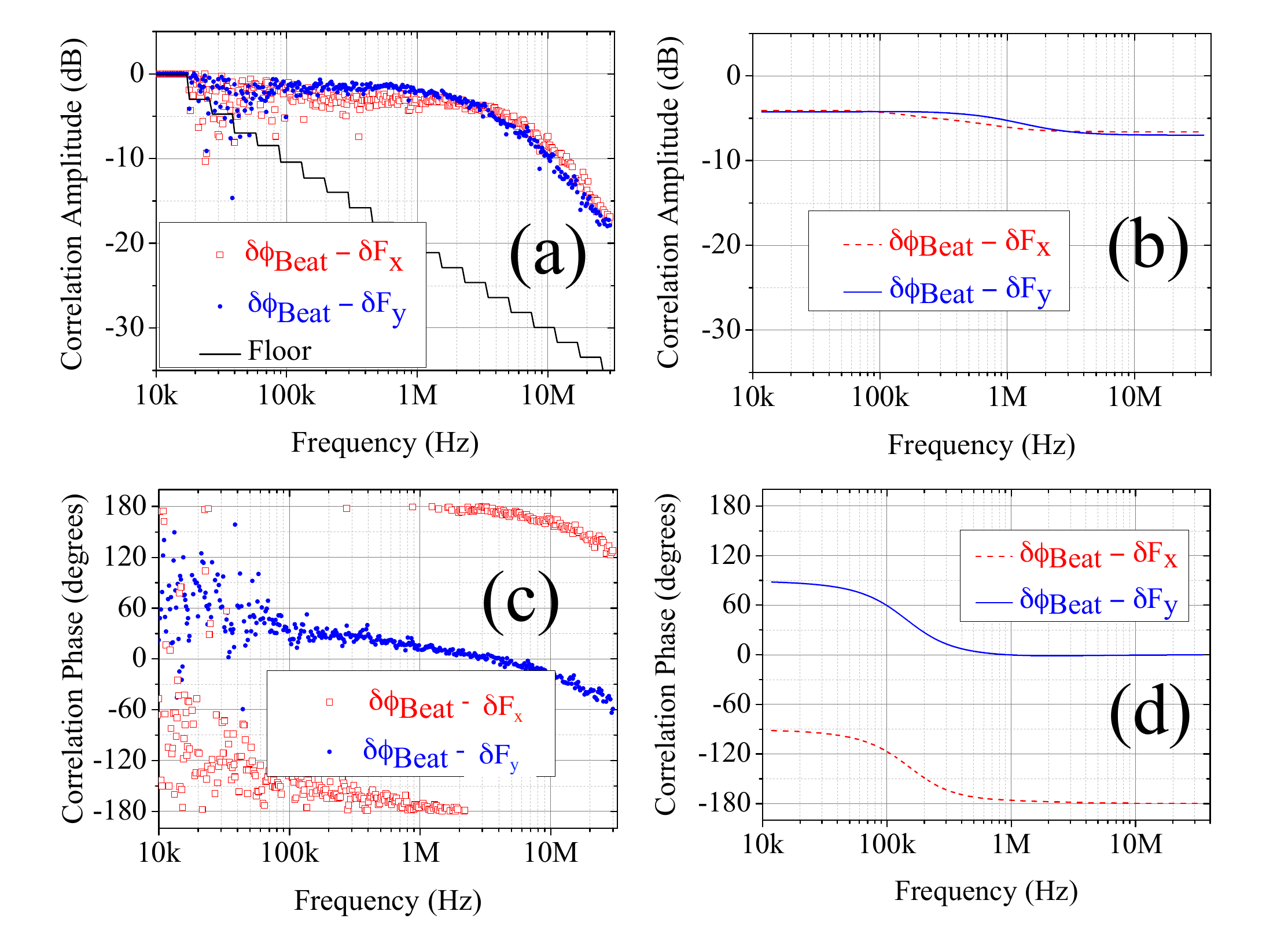}
\caption{(Color online) Correlation of RF phase noise with the intensity noises for weak coupling. Correlation amplitude: (a) experiment, (b) theory; correlation phase: (c) experiment, (d) theory.}
\label{fig:8}
\end{figure}

The measurements reproduced in the preceding subsections, dealing with intensity noise spectra and correlation spectra on the one hand, and phase noise spectra on the other hand, have shown very good agreement with the model of Sec.\ \ref{sec:Theory}. However, in order to further test this model and in particular the scenario that we have proposed for the propagation on the pump noise to the RF beatnote phase noise, we have chosen to measure the correlation between the phase noise of the RF beatnote and the intensity noises of the two laser modes generating the RF beatnote for both weak and moderately strong coupling, and to compare them with the theoretically expected spectra. The experimental and theoretical results for weak coupling are reproduced in Fig.\ \ref{fig:8}. 
The measured correlation amplitude spectra are reproduced in Fig.\ \ref{fig:8}(a). The red open squares and blue filled circles represent the amplitude of correlation between the RF phase noise with the intensity noises of x- and y-polarized modes, respectively. The correlation amplitude for the two modes are not perfectly identical as their intensity noises are different (see Figs.\ \ref{fig:4}(a),(b)) due to the difference between their losses and/or gains. The correlation amplitude is high ($-4$~dB) for low frequencies and it starts to decrease for high frequencies starting from 1 MHz. The theoretical model satisfactorily reproduces  this behavior, as shown in Fig.\ \ref{fig:8}(b). In Fig.\ \ref{fig:8}(a), the decrease of the correlation amplitude for frequencies higher than 5 MHz is due to the fact that the RF phase noise reaches the measurement floor level for these frequencies (Fig.\ \ref{fig:7}(a)). This explains the discrepancy here with theory (Fig.\ \ref{fig:8}(b)). The experimental correlation phase spectra are represented by Fig.\ \ref{fig:8}(c). The corresponding theoretical spectra (Fig.\ \ref{fig:8}(d)) show good agreement with the experiment. The correlation phase is $ + \pi/2 $ ($ -\pi/2 $) for frequencies lower than 100 kHz, then it starts to roll off and reaches zero ($ \pi $) at about 1 MHz and remains at zero ($ \pi $) for all higher frequencies for the y-polarized (x-polarized) mode. The correlation phases for x- and y-polarized mode differ by $ \pi $ for all the frequencies, due to the minus sign in Eq.\,(\ref{eqfb03}). 
  
\begin{figure}[htbp]
\centering
\includegraphics[width=1.0\columnwidth]{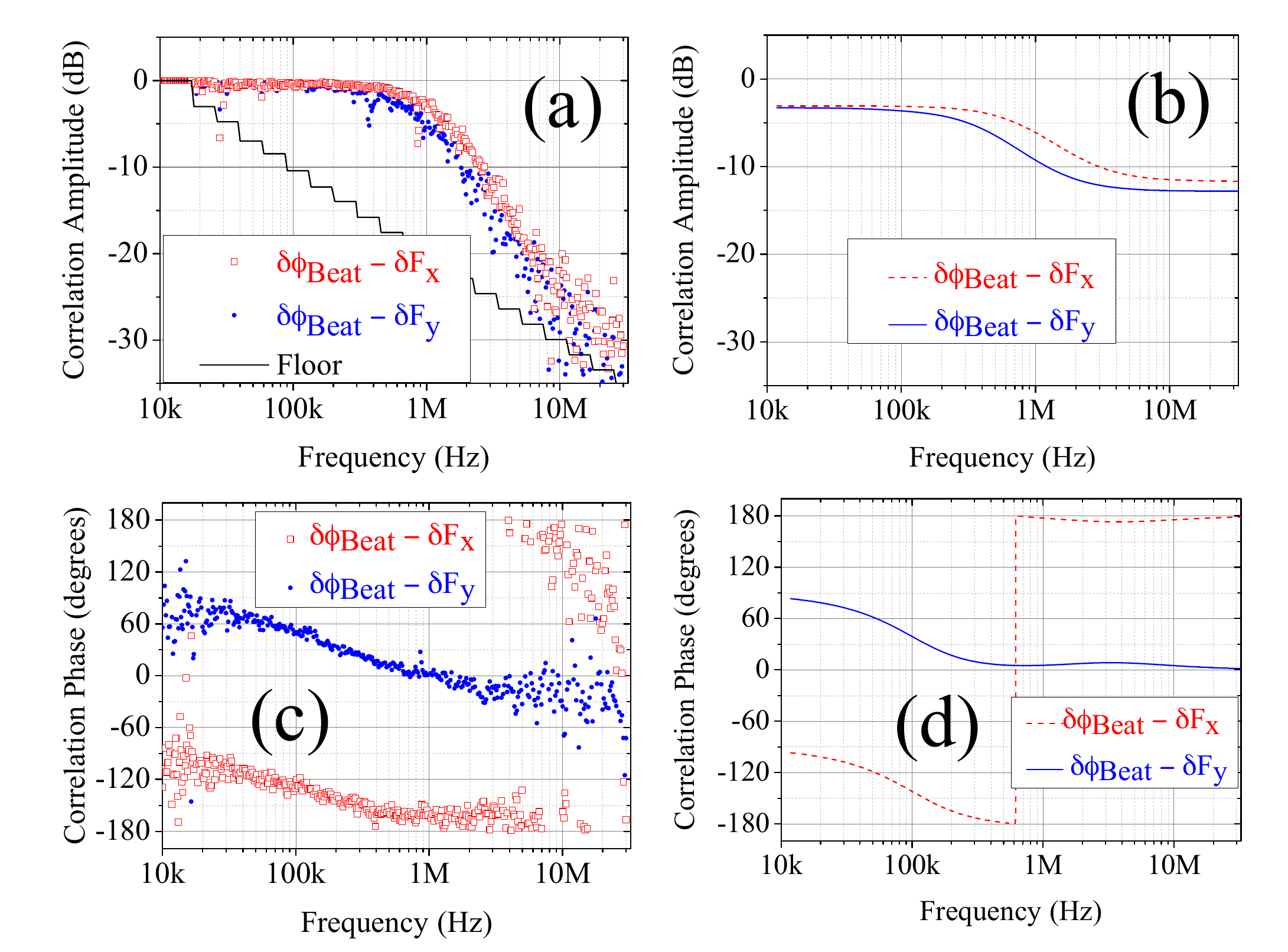}
\caption{(Color online) Correlation of RF phase noise with the intensity noises for moderately strong coupling. Correlation amplitude: (a) experiment, (b) theory; correlation phase: (c) experiment, (d) theory. }
\label{fig:9}
\end{figure}
 The results in the case of moderately strong coupling are reproduced in Fig.\ \ref{fig:9}. Here again we observe a good agreement between experiment (Figs.\ \ref{fig:9}(a),(c)) and theory (Figs.\ \ref{fig:9}(b),(d)). In this case also, the difference of correlation amplitudes for the two modes (red open squares and blue filled circles) is related to their different intensity noises (see Figs.\ \ref{fig:5}(a),(b)) due to the nonidentical losses and/or gains. For moderately strong coupling, as shown in Figs.\ \ref{fig:9}(a),(b), the correlation amplitude is higher ($- 2$~dB) compared to weak coupling for low frequencies. But again the correlation amplitude starts to decrease for high frequencies starting from 1 MHz, like in case of weak coupling. Here also, the decrease of correlation amplitude for frequencies higher than 5 MHz is due to the fact that the phase noise reaches the measurement noise floor for these frequencies (see Fig.\ \ref{fig:7}(a)). Additionally, the correlation phases associated with the two modes (see Figs.\ \ref{fig:9}(c),(d)) always differ by $ \pi $, like in the weak coupling situation. Moreover, for frequencies lower than 100 kHz, the correlation phase is $ + \pi/2 $ (resp. $ -\pi/2 $), then the rolling off starts and it becomes zero (resp. $\pm \pi $) at about 1 MHz and remains at zero (resp. $\pm \pi $) for all higher frequencies for the $y-$polarized (resp. $x-$polarized) mode. 
 
 For both weak and moderately strong coupling, the change of correlation phase by $ \pi/2 $ from $10\;\mathrm{kHz}$ to few hundreds of kHz for both $x-$ and $y-$polarized modes is related to the low-pass filter behavior of the thermal phase noise (Fig.\ \ref{fig:7}(a)). Indeed,  a $ \pi/2 $-phase change is associated with this passage through the cut-off frequency, like for any first-order filter. Finally, we observe that the correlation amplitude for frequencies lower than 1 MHz is stronger (- 2~dB) for moderately strong coupling than for weak coupling (- 4 dB), due to the fact that the pump noise correlation factor $ \eta $ is stronger for moderately strong coupling than for weak coupling (see Fig.\,\ref{fig:3}(a)). 

 \smallskip
\section{conclusion}
In this paper, we have reported the spectral behavior of the correlation between the different noises in a two orthogonal polarization dual-frequency VECSEL operating at telecom wavelength. Specifically, we have investigated the spectra of correlation between intensity noises of the two linear-orthogonal polarizations, the correlation between the phase noise of the RF beatnote and the intensity noises of the two eigenpolarizations. The effect of laser dynamics, which is governed by the strength of the nonlinear coupling between the modes, on the spectral behavior of noise correlations has been experimentally and theoretically investigated. We have found that the spectral behavior of the pump noise as well as the correlation between the pump noises entering  the two spatially separated laser modes play a major role in the amplitudes and phases of the correlations between the different noises of our DF-VECSEL. A modified rate equation model has exhibited excellent agreement with the experimental results. Moreover, it has allowed to interpret the correlation results in terms of linear response of two coupled over damped oscillators. This noise correlation study enables us to reach an understanding of the different mechanisms deteriorating the noise performance of a DF-VECSEL at telecom wavelength and hence in future will help us to improve the noise performance required for different applications. This understanding would even enable in near future to shape the noise spectra of the laser according to the targeted application.

\section*{Acknowledgments}
The authors thank Gr\'egoire Pillet for providing the computer codes to compute the correlation spectra and Lo\"ic Morvan for his help. This work was partially supported by the Agence Nationale de la Recherche (ANR) (NATIF, ANR-09-NANO-012-01).

\end{document}